\documentstyle[11pt,aaspp,psfig]{article}
\begin{document}

\title{New Modeling of the Lensing Galaxy and Cluster of Q0957+561: 
Implications for the Global Value of the Hubble Constant}

\author{Kyu-Hyun Chae}
\affil{Department of Physics \& Astronomy, University of Pittsburgh,
Pittsburgh, PA 15260}

\begin{abstract}
The gravitational lens 0957+561 is modeled utilizing recent observations of
the galaxy and the cluster as well as previous VLBI radio data which 
have been re-analyzed recently. The galaxy is modeled by a power-law 
elliptical mass density with a small core while the cluster is modeled 
by a non-singular power-law sphere as indicated by recent observations. 
Using all of the current available data, the best-fit model has  
$\chi^2_{\mbox{\scriptsize min}}$/N$_{\mbox{\scriptsize dof}}$ $\approx$ 
6 where the $\chi^2$ value is dominated by a small portion of the 
observational constraints used; this value of the reduced $\chi^2$ is 
similar to that of the recent FGSE best-fit model by Barkana et al. 
However, the derived value of the Hubble constant is significantly different 
from the value derived from the FGSE model. We find that the 
value of the Hubble constant is given by $H_0$ = $69^{+18}_{-12}(1-\kappa)$ 
and $74^{+18}_{-17}(1-\kappa)$ km s$^{-1}$ Mpc$^{-1}$ with and without 
a constraint on the cluster's mass, respectively, where $\kappa$ is the 
convergence of the cluster at the position of the galaxy and the range for 
each value is defined by 
$\Delta\chi^2 = \chi^2_{\mbox{\scriptsize min}}$/N$_{\mbox{\scriptsize dof}}$.
Presently, the best achievable fit for this system is 
not as good as for PG 1115+080, which also has recently been used 
to constrain the Hubble constant, and the degeneracy is large. 
Possibilities for improving the fit and reducing the degeneracy are discussed.
\end{abstract}


\section{Introduction}
The gravitationally lensed ``double quasar'' Q0957+561 is 
the first lensed system for which time delay between image components 
has been measured with considerable confidence and accuracy 
(Haarsma et al.\ 1999; Pelt et al.\ 1998; Kundi\'{c} et al.\ 1997; 
Pijpers 1997; Oscoz et al.\ 1997; Schild \& Thomson 1997). In principle, 
the measured time delay ($417\pm 3$ d, Kundi\'{c} et al.\ 1997) can be 
turned into an accurate determination of the Hubble constant, $H_0$,
using the method outlined by Refsdal (1964, 1966; see Schneider, Ehlers, 
\& Falco 1992 for a pedagogical introduction).
For this, the potential of the lens (i.e.\ its mass distribution) 
must be well-determined consistent with observational constraints. 
Many important observational constraints on Q0957+561 have come from the
VLBI observations of the radio jets and cores of the two image components 
(Garrett et al.\ 1994; Gorenstein et al.\ 1988; Gorenstein et al.\ 1984). 
The Garrett et al.\ (1994) data have been re-analyzed just recently and 
improved constraints are given by Barkana et al.\ (1999, hereafter Ba99).

The lensing of Q0957+561 is due to the combined effect of a massive elliptical
galaxy and a cluster at redshift $z=0.36$ of which the elliptical galaxy is 
the brightest member. The cluster has been studied using optical observations 
(Angonin-Willaime, Soucail, \& Vanderriest 1994; Garrett, Walsh, \& Carswell 
1992), X-ray observations (Chartas et al.\ 1998; Chartas et al.\ 1995), and 
weak lensing effects (Fischer et al.\ 1997; Dahle, Maddox, \& Lilje 1994). 
These studies indicate that the lensing galaxy is positioned close to the 
center of the cluster, and the cluster's convergence in the region of the 
image is significant. However, the cluster's mass distribution could not be 
accurately determined from those studies. The lensing galaxy has been detected
both optically (Bernstein et al.\ 1997, hereafter Be97; Stockton 1980) 
and in the radio (Roberts et al.\ 1985; Gorenstein et al.\ 1983). 
The more recent HST observations by Be97 appear to have significantly
reduced the uncertainty in the position of the galaxy. 
The HST optical position (G1) is consistent with the VLBI position 
(G$'$), but is inconsistent with the VLA position (G). The Be97 
observations also reveal new possible lensed features of 
two blobs and two knots along with an arc in the system. 

Present observational constraints (the VLBI and HST constraints) now
clearly exclude both the Grogin \& Narayan (1996) softened power-law sphere 
(SPLS) model and the Falco, Gorenstein, \& Shapiro (1991) model (FGS) of 
a King profile sphere with a central black hole, which were previously 
the best-fit models. Ba99 extended the SPLS and FGS models to include an 
arbitrary ellipticity to the galaxy, and the new versions of models [softened
power-law elliptical mass distribution (SPEMD) and FGSE] fit the observations  
better. According to their study, the FGSE model gives a somewhat 
better fit than the SPEMD model [$\bar{\chi}^2_{\mbox{\scriptsize min}}$
($\equiv$ $\chi^2_{\mbox{\scriptsize min}}$/N$_{\mbox{\scriptsize dof}}$) 
$\approx$ 6 vs. 10]. Notably, the value of $H_0$ derived from the FGSE model 
is $\gtrsim$ 100 km s$^{-1}$ Mpc$^{-1}$. In previous models of Q0957+561, 
the lensing cluster has been  modeled by a quadrupole term (e.g.\ Ba99; 
Grogin \& Narayan 1996; Falco, Gorenstein, \& Shapiro 1991) because of 
its simplicity and assuming that higher order terms can be neglected in 
the expansion of the cluster potential. In addition, Ba99 considered a 
singular isothermal sphere (SIS) to model the cluster for the FGSE model of 
the galaxy while Bernstein \& Fischer (1999) considered higher
order terms in the cluster expansion.  
Ba99 found that for the FGSE model of the galaxy, a SIS model of the cluster
did not lead either to an improved $\bar{\chi}^2_{\mbox{\scriptsize min}}$ or
a significantly different value of $H_0$ compared with the case using a
quadrupole term. Thus, for the FGSE model of the galaxy, it appears 
that consideration of a realistic cluster mass model is not crucial. 
However, the quadrupole term alone may not accurately account for the cluster
contribution, especially since recent observations indicate the cluster's mass 
center can be very close to the galaxy (Fischer et al.\ 1997; 
Angonin-Willaime et al.\ 1994), and there is no {\it a priori} reason that 
we need only consider the SIS model for the cluster. Bernstein \& Fischer's 
(1999) consideration of higher-order terms can be a better approximation of 
the cluster potential than the single quadrupole term, however, the low 
$\chi^2$ values of their best-fit models depended, in large part, on their 
selective use of the current observational constraints (see \S 5).

In this study we consider a more general model of the 
cluster, i.e., a power-law sphere while modeling the galaxy 
with a power-law elliptical mass density with a small core [i.e.\ the
SPEMD model with a fixed small core (see \S 2 and \S 3)]. We also consider 
the case where the cluster is elliptical. The reduced $\chi^2$ of the best-fit
model of this type is similar to that of the best-fit FGSE model. 
However, the derived value of $H_0$ is significantly different
from the value derived from the FGSE model.  It turns out that neither the
core radius nor the power-law radial index of the cluster can be determined
from fitting the model to the current observational constraints. This is
because the variation of either the radial index or the core radius 
(within a wide range) does not significantly alter the fit,
while the convergence of the cluster ($\kappa$) at the position of the galaxy 
(and thus the derived value of $H_0$) varies. Thus, we can only 
derive a scaled [by a factor of $(1-\kappa)$] value of $H_0$. 
The outline of this paper is as follows. We first review the adopted
observational constraints (\S 2). In \S 3, we present the lens model  
and review its parameters. In \S 4, we present the results of fitting the 
model to the observational constraints, and derive a value for $H_0$. 
In \S 5, we discuss possible future improvements to the determination 
of $H_0$ using the Q0957+561 lens.

\section{Summary of Current Observational Constraints}
The optical image of Q0957+561 consists of two point-like quasars A and B 
$\approx 6''$ apart on the sky. This is only a weak constraint on a lens 
model. However, the radio image of Q0957+561 provides many strong constraints.
In particular, the extended radio jets along with the radio cores can be 
used to derive both accurate relative positions of the jets with respect to 
their cores and between the cores, and relative magnification matrix between 
the corresponding components of A and B. 
The separation between the cores was measured 
with an uncertainty of $0''.00004$ using VLBI observations (Gorenstein et al.\ 
1984; Falco et al.\ 1991). Further VLBI observations at $\lambda = 13$ cm 
by Gorenstein et al.\ (1988) identified three jets in each quasar, which
were used to derive a relative magnification matrix [{\bf M}$_{BA}$] = 
[$\partial {\bf x}_B / \partial {\bf x}_A$] between quasars A and B. 
More recently, VLBI observations at $\lambda = 18$ cm by Garrett et al.\ 
(1994) separated the jets into five components for each quasar, from which
they derived an improved relative magnification matrix along with its gradient.
However, Ba99 employed improved analysis procedures to re-analyze the raw data
of Garrett et al.\ (1994), in particular, they corrected some significant 
errors in the VLBI data reduction packages adopted by Garrett et al.\ (1994)
and fitted the jet component parameters (i.e.\ positions, fluxes, and Gaussian 
model parameters) and the magnification transformation parameters 
simultaneously in a single step, and so we will use the constraints given 
by Ba99 in this study. 

Radio observations of the system have resulted in the detection of radio 
sources near the optical center of the lensing galaxy (Roberts et al.\ 1985; 
Gorenstein et al.\ 1983). While both the VLA position G (Roberts et al.\ 1985)
and the VLBI position G$'$ (Gorenstein et al.\ 1983) were reported to have an 
uncertainty of 1 mas, they differed $30\sigma$ from each other. The optical 
position (G1) of the lensing galaxy as measured by Stockton (1980) is more 
consistent with the VLBI position, however, his measurement uncertainty is 
large (30 mas). Thus, in previous studies which modeled the lens, it was not 
clear which observed position to use as a constraint. However, HST observations
of the system by Be97 were used to determine the optical position of the 
lensing galaxy with an uncertainty of 3.5 mas. The HST position is consistent 
with the VLBI position within 10 mas, but it is clearly inconsistent with the 
VLA position. Whether the HST position or the VLBI position are used to 
constrain the lens model (\S 4) makes little difference; however we will use 
the VLBI position. Another constraint to be used is the core flux ratio B/A 
derived from VLBI/VLA observations and observations of the optical emission 
lines (Conner, Leh\'{a}r, \& Burke 1992; Schild \& Smith 1991). The above 
constraints, derived mostly from VLBI observations, are used as our ``primary
constraints'' [we mean by this that these constraints have been derived from 
confirmed lensed features, and that they are used in the first step of the
fitting procedure (see \S3)] on the lens model; 
they are summarized in Table 1. 

The Be97 HST observations detected possible lensed images of other background 
sources. Blobs 2 \& 3 are probably images of a common background galaxy. 
Knots 1 \& 2, which form an arc, are also probably images of a common source. 
While a lens model which is based on the primary constraints easily 
satisfies the constraints from the knots, the constraints from the blobs
turn out to be useful for distinguishing between models. The constraints 
from the blobs and knots are used as ``secondary constraints'' which are
summarized in Table 4. The total number of primary and secondary constraints 
is 25. All of these constraints are used to define the ``goodness of fit'', 
$\chi^2$, taking into account correlation coefficients for some constraints 
(Table 2 and Table 3).

Another observational constraint is the upper limit on the relative brightness 
of any third image (image ``C'') which is predicted by a lens model that 
has a smooth mass distribution at the central region of the galaxy. The VLBI 
observations by Gorenstein et al.\ (1984) put a $5\sigma$ upper limit of C/B 
= 3.3\% while the optical observations by Stockton (1980) found C/B $<$ 2\%. 
The relative brightness of the third image is mostly related to the 
mass distribution at the central region of the lens, which is effectively
described by a core radius parameter. The observed low values of the upper 
limit on the relative brightness of a third image implies that the core radius 
of the lens model is very small (see \S 3). If the core radius is sufficiently
small, then the lensing properties of the model become insensitive to
the value of the core radius, since mass distributions with different core 
radii (and different central densities) are essentially the same outside 
the small central regions. Thus, we can fix the core radius at a small value 
which ensures that the relative magnification of the third image predicted 
by the model is always less than the observational upper limit. This approach
is employed in this study.

\section{An Observationally Motivated Model of the Lens}
We consider the following forms for the mass distributions of 
the lensing galaxy
\begin{equation}
\Sigma_{\mbox{\scriptsize gal}}(r,\theta) = \frac{\Sigma_{1}}
   {\left\{1+\left(\frac{r}{r_{1}}\right)^2
  [1+e_1 \cos 2(\theta-\theta_{1})] \right\}^{(\nu_1-1)/2}}
\end{equation}
and cluster
\begin{equation}
\Sigma_{\mbox{\scriptsize cl}}(r') = 
\frac{\Sigma_2}{\left\{1+\left(\frac{r'}{r_2}\right)^2 \right\}^{(\nu_2-1)/2}}.
\end{equation}
Here  $\Sigma_{i}$ ($i=1,2$) are the central surface mass densities, 
$r_i$ are the core radii, and  $\nu_i$ are the radial indices. 
Radial index $\nu = 1$, 2, and 3 correspond to radial profiles of constant 
surface density, isothermal, and modified Hubble law, respectively, so
that it is related to the parameter $\eta$ used by Grogin \& Narayan (1996)
and Ba99 via $\nu = 3 - \eta$.
The parameter $e_1$ ($>0$) is related to the galaxy's
ellipticity $\epsilon_1$ ($\equiv 1-b/a$ where $a$, $b$ are the major,
minor axes respectively) via $ \epsilon_1 = 1-[(1-e_1)/(1+e_1)]^{1/2}$.
The parameter $\theta_1$ is the position angle (north through east) of the 
galaxy. Primed coordinates are used for the cluster's mass 
distribution since the cluster's center is off from the galaxy's center.
The relative position of the cluster with respect to the galaxy
is represented by the distance between them, $d_{12}$, and the position
angle of the cluster's center as viewed from the galaxy, PA$_{12}$.
The above model will be referred to as ``PEM+PS'' (meaning Power-law 
Elliptical Mass + Power-law Sphere) model.

Throughout we assume a standard cosmology with a cosmological matter-energy 
parameter $\Omega_{\mbox{\scriptsize Matter}} = 1$ and a cosmological 
vacuum-energy parameter $\Omega_{\Lambda} = 0$, and we use the usual 
definition $h = H_0/ 100$ km s$^{-1}$ Mpc$^{-1}$. For this choice of cosmology
we have $1'' \approx 3.0 h^{-1}$ kpc at the redshift of the lens. 
For an open universe with $\Omega_{\mbox{\scriptsize Matter}} = 0.3$ (and
$\Omega_{\Lambda} = 0$), an estimate of $h$ increases by $\approx 7\%$.
As explained in \S 2, the present observational constraint on the relative 
brightness of a third image implies that the core radius of the galaxy, 
$r_1$, should be very small. For example, the upper limit on the core radius
of a galaxy with radial index $\nu_1 = 1.72$ in equation (1) is 
$r_1 \approx 10 h^{-1}$ pc for C/B $\leq$ 3.3\%.
We fix the core radius at $r_1 = 0.1 h^{-1}$ pc, 
which is small enough to satisfy the observational constraint for any value 
of $\nu_1$, but is an arbitrary choice otherwise (see \S 2).
The other parameters of the galaxy (i.e.\ central density, ellipticity, and
position angle) are free parameters for the model. Thus, our model of the
lensing galaxy can be considered as a dark matter dominated mass model. It will
be interesting to compare the determined mass distribution of the lensing 
galaxy with its observed light distribution (Be97).

Present observational studies of the lensing cluster indicate that the 
cluster's mass distribution is more consistent with an extended core 
(Chartas et al.\ 1998; Fischer et al.\ 1997), although the uncertainties in 
the derived values are too large for them to be directly used to constrain
the lens model. 
Thus, we fix the core radius at an arbitrary large value (e.g.\ several 
arcseconds) and derive a value of $h$ with the fixed core radius. And then, 
we study how the derived value of $h$ is affected if we vary the core radius
from the chosen value. There is no observational constraint on the radial
index of the cluster. Thus, we fix the radial index at an arbitrary value,
e.g., $\nu_2 = 2$ (isothermal profile), and then study how the derived value 
of $h$ is affected when $\nu_2$ is varied.

To calculate the lensing properties (i.e.\ deflection, magnification, and 
relative light travel time) of the mass distribution implied by
equation (1) we use the series method by Chae, Khersonsky, \& Turnshek (1998);
for the details, the reader is referred to the paper and references therein.
In fitting the observational constraints of Q0957+561, we must use 
an accurate and robust method of calculation since the fractional errors of 
the VLBI data are as small as $\sim 10^{-5}$. The calculational accuracy
of the series method can be made orders of magnitude smaller than the 
observational errors by controlling the truncation of the series. This was  
tested for the isothermal case of $\nu = 2$ using the analytic solution of 
Kormann, Schneider, \& Bartelmann (1994).\footnote{The calculational accuracy
of the series method depends on the value of the core radius. For our chosen
value of $r_c = 10^{-4} h^{-1}$ kpc, the fractional calculational errors are
$\sim 10^{-9}$ to $10^{-7}$ for $\epsilon \lesssim 0.5$.}  
Since the series converge well for an arbitrary value of $\nu$, 
we can be confident that the series calculation
gives sufficiently accurate results. Given that the total number of constraints
is 25 and the total number of free parameters [including the eight
coordinates of the four sources (i.e.\ core, jet, blob and knot) and two
redshifts of the blob and knot] is 17, the minimization of the $\chi^2$
is not a simple numerical task. For this reason, we first fit the model to
the primary constraints only, so that we can work with smaller numbers of
constraints and free parameters. Once best-fit values based on the primary 
constraints are determined, the secondary constraints and the remaining 
parameters are considered, and then the entire set of parameters are adjusted 
to minimize the $\chi^2$. 

\section{Results of Fitting the Model to the Observational Constraints:
Bounds on the Hubble Constant}
Because of the degeneracies in the cluster's core radius 
($r_2$) and radial index ($\nu_2$), we first consider 
fixed values of $r_2$ and $\nu_2$, and then study how varying them affects
the derivation of $h$. We set $\nu_2 = 2$ and $r_2 = 15 h^{-1}$ kpc 
($\approx 5''$). We find that the model has significant degeneracies in
parameters $\nu_1$ (the galaxy's radial index) and $d_{12}$ (the separation
between the cluster and the galaxy). Thus, we fix the parameters $\nu_1$
and $d_{12}$ for each model and, by incrementing them in steps of 0.02 and
$1''$ respectively, obtain a grid of models in the 2-dimensional parameter 
space spanned by $\nu_1$ and $d_{12}$. A model with $\nu_1 = 1.72$ and 
$d_{12} = 9''$ has the lowest value of $\chi^2$ of 
$\chi^2_{\mbox{\scriptsize min}} \approx 49.6$ with 8 degrees of freedom
($\bar{\chi}^2_{\mbox{\scriptsize min}} \approx 6.2$). 
As the parameters $\nu_1$ and $d_{12}$ are varied from the best-fit 
values, $\chi^2$ increases moderately. A confidence limit (CL) ellipse 
defined by $\Delta \chi^2 = \bar{\chi}^2_{\mbox{\scriptsize min}}$ is shown
in Figure 1 (full line). Although this results in a larger region of parameter
space than the conventional $1\sigma$ (i.e.\ 68\% CL) region, 
we adopt it as in Ba99 and Grogin \& Narayan (1996) 
since the best-fit model has a poor overall $\chi^2$.
 
For the best-fit model ($\nu_1 = 1.72$, $d_{12} = 9''$), 
we find $h = 0.54$ which is obtained from
$h = \Delta\tau_{\mbox{\scriptsize AB}}^{\mbox{\scriptsize (model)}}/
\Delta\tau_{\mbox{\scriptsize AB}}^{\mbox{\scriptsize (observed)}}$ 
= (225 d)/(417 d) = 0.54, where 
$\Delta\tau_{\mbox{\scriptsize AB}}^{\mbox{\scriptsize (model)}}$ is 
calculated for $H_0 = 100$ km s$^{-1}$ Mpc$^{-1}$. 
In the $\Delta \chi^2 = \bar{\chi}^2_{\mbox{\scriptsize min}}$ range, 
$0.39 < h < 0.70$. However, this value of $h$ was obtained for the fixed 
cluster parameters of $\nu_2 = 2$ and $r_2 = 15 h^{-1}$ kpc. Although these 
choices are the best values obtained by Fischer et al.\ (1997), their allowed 
range of $r_2$ is large, and there is no {\it a priori} reason that the 
cluster's mass distribution should have an exact isothermal profile. 
Thus, we must study how the above derived value of $h$ is affected as 
the cluster parameters are varied. The effect of varying the cluster's 
radial index turns out to be trivial; for a profile shallower 
($\nu_2 < 2$) or steeper ($\nu_2 > 2$) than the isothermal, the fit results 
are virtually the same except that the cluster's convergence at the position 
of the galaxy ($\kappa$), i.e.\ at $r=0$, is altered and the value of $h$ 
is scaled by a factor of the ratio of the values of ($1-\kappa$). 
In Figure 2 are shown the values of ($\chi^2$, $h/(1-\kappa)$, $\kappa$)
for various example cluster parameter sets. We see that for a wide range of 
$1.5 \leq \nu_2 \leq 2.5$ around the isothermal value ($\nu_2 = 2$) 
at core radius $r_2 = 15 h^{-1}$ kpc ($\approx 5''$), 
$\Delta\chi^2 < 0.12 \bar{\chi}^2_{\mbox{\scriptsize min}}$ and 
$h/(1-\kappa)$ changes by no larger than 1.4\% while $\Delta\kappa > 0.22$. 
The effect of varying the cluster's core radius can be more complicated. 
Nevertheless, we find that for the observational range of $r_2$ by Fischer 
et al.\ (1997), the effect is simple (see Fig.\ 2). Namely, 
when the core radius is increased or decreased from $r_2 \approx 5''$ 
for the isothermal profile, the $\chi^2$ varies little 
($\Delta\chi^2 < 0.18 \bar{\chi}^2_{\mbox{\scriptsize min}}$), 
the value of $\kappa$ is altered, and the value of $h$ scales, to a good 
approximation (up to an error of 2.8\%), according to ($1-\kappa$) for 
$0 \lesssim r_2 \lesssim 10''$. In summary, for a range of the cluster's 
radial index and core radius ($1.5 \leq \nu_2 \leq 2.5$, 
$0 \lesssim r_2 \lesssim 10''$) which we investigated with the current 
observational constraints, we find that there is no significant change of 
$\chi^2$ ($\Delta\chi^2 < 0.26 \bar{\chi}^2_{\mbox{\scriptsize min}}$),
and that the change of $h$ is dictated by $(1-\kappa)$ to an error no
larger than $\approx 4$\%. Therefore, the mass-sheet degeneracy 
(Falco, Gorenstein, \& Shapiro 1985) appears to persist even when we
use a full potential model of the cluster. (We remind ourselves that when 
the quadrupole approximation was adopted in previous lens models of Q0957+561, 
this degeneracy was exact.)

However, for each radial profile of the cluster, 
the $\chi^2$ has a ``minimum'' value 
at a certain finite core radius (e.g.\ at $r_2 \approx 2''$ for 
the isothermal case) and the $\chi^2$ increases consistently as the core 
radius is varied, although we concluded above that the variation of the 
$\chi^2$ was insignificant with the current observational constraints. 
There is a subtlety here. Even though the total $\chi^2$ remains virtually 
invariant as the core radius is varied for a fixed radial index, 
the $\chi^2$ contributions due to the primary constraints 
and those due to the secondary constraints vary significantly.
For instance, for the isothermal profile ($\nu_2 = 2$), the model with 
$r_2 \approx 0''$ has $\chi^2_{\mbox{\scriptsize p}} \approx 35.6$ and
$\chi^2_{\mbox{\scriptsize s}} \approx 14.0$ while the model with
$r_2 \approx 10''$ has $\chi^2_{\mbox{\scriptsize p}} \approx 40.2$ and
$\chi^2_{\mbox{\scriptsize s}} \approx 10.2$, where 
$\chi^2_{\mbox{\scriptsize p}}$ and $\chi^2_{\mbox{\scriptsize s}}$ denote 
the $\chi^2$ contributions due to the primary and secondary constraints,
respectively. As will be seen in \S 5, the radio jet positions have the
largest $\chi^2$ contribution out of the the primary constraints, 
so does the blob magnification ratio out of the secondary constraints. 
This suggests that better observational knowledge 
of the radio jets and the blob magnification ratio could break the present 
``superficial'' degeneracy of the core radius for a given value of the
radial index. When the radial index is varied for a fixed value of the
core radius, the $\chi^2_{\mbox{\scriptsize p}}$ and 
$\chi^2_{\mbox{\scriptsize s}}$ each change by only $\approx 1$ for 
$\Delta\nu = 1$. This means that the degeneracy of the cluster's radial 
index is much more significant with the present observational constraints. 
Unless new independent observational constraints are revealed, the degeneracy 
of the cluster's radial index could not be broken. 

As seen in the above, the degeneracy of the cluster's core radius is weaker 
than that of the cluster's radial index. However, the determination of
$h/(1-\kappa)$ is not significantly affected by varying the core radius.
To illustrate this, CL ellipses are drawn in Fig.\ 1 for two other values of 
$r_2$, i.e., $r_2 \approx 3''$ (dashed line) and $r_2 \approx 7''$ 
(dotted line). As seen in the figure, confidence region of the parameter 
space is altered only slightly whether as the core radius is increased or
decreased. The $\Delta \chi^2 = \bar{\chi}^2_{\mbox{\scriptsize min}}$ range 
of $h/(1-\kappa)$ is nearly unchanged for $0 \lesssim r_2 \lesssim 10''$. 
From the grid of models with $\nu_2 = 2$ and $r_2 = 15 h^{-1}$ kpc 
($\approx 5''$), we find $h = 0.74^{+0.18}_{-0.17} (1-\kappa)$.

As we considered a CL by 
$\Delta \chi^2 = \bar{\chi}^2_{\mbox{\scriptsize min}}$ in the above,
we could consider a CL by 
$\Delta \chi^2 = 4 \bar{\chi}^2_{\mbox{\scriptsize min}}$.
However, the confidence region within so defined a CL is too 
large to be useful to constrain $h$. If the cluster's mass is constrained, the
ranges of the parameter space can be reduced. Unfortunately, the cluster's
mass is not well-determined from present observations.
Figure 3 shows confidence regions of the parameter space of the model
with $\nu_2 = 2$ and $r_2=15 h^{-1}$ kpc including the cluster's mass as 
a constraint determined directly from weak lensing effects by Fischer et al.\ 
(1997; $\Sigma_0 = 0.36\pm 0.11 h \times 10^{10}$ M$_{\odot}$ kpc$^{-2}$
for $\nu = 2$ and $r_c \approx 5''$).\footnote{Here we do not constrain the
cluster's positions, and thus we do not constrain the cluster's convergence
at $r=0$ although we do constrain its mass.} 
The parameter space is slightly better constrained compared 
with the case that the cluster's mass is not constrained 
(Fig.\ 1). We find $\bar{\chi}^2_{\mbox{\scriptsize min}}\approx 5.6$ for
$\nu_1 = 1.70$ and $d_{12} = 10''$. The derived value of the Hubble constant 
is $h = 0.69^{+0.18}_{-0.12} (1-\kappa)$ with the cluster's mass 
constrained. The measured velocity dispersions of the lensing galaxy
(Tonry \& Franx 1998; Falco et al.\ 1997; Rhee 1991) can, in principle,
be used to constrain the mass of the galaxy, which can be useful for
reducing the degeneracy in the parameter space of the PEM+PS model. Stellar 
orbit modeling methods have been applied to the best-fit SPLS mass profile of 
Grogin \& Narayan (1996) to infer the mass of the galaxy 
(Romanowsky \& Kochanek 1999). Similar studies in the future for the PEM
galaxy model could be useful.

Table 5 summarizes the values of the parameters for the model with $\nu_2 = 2$ 
and $r_2 = 15 h^{-1}$ kpc with the cluster's mass unconstrained. 
The theoretical images of the radio core and the brightest radio jet
for the best-fit model are shown in Figure 4. Figure 5 shows theoretical
images of the optical blob and knot along with an arc. The model 
galaxy's radial profile is somewhat shallower than the isothermal profile 
$\nu = 2$, i.e., the mass within radius $r$ increases faster than the 
isothermal case as $r$ increases. The model galaxy's ellipticity 
($\epsilon_{\mbox{\scriptsize model}} \approx 0.01$ to 0.38) is similar to 
the ellipticity of the observed light ($\epsilon_{\mbox{\scriptsize light}} 
\approx 0.05$ to 0.49; Be97). The observed position angle of the light 
distribution is PA$_{\mbox{\scriptsize light}} \approx 32\deg$ to $68\deg$ (Be97). 
The position angle of the best-fit model (PA$_{\mbox{\scriptsize model}} = 64\deg$)
is in agreement with the light distribution. However, in the 
$\Delta \chi^2 = \bar{\chi}^2_{\mbox{\scriptsize min}}$ range,
$6\deg \lesssim$ PA$_{\mbox{\scriptsize model}} \lesssim 166\deg$.
Here it is interesting to note that for the majority of 17 gravitational
lens galaxies studied by Keeton, Kochanek, \& Falco (1998), the position angles
of the light and the mass (i.e.\ lens model) are the same to $\lesssim 10\deg$.
In other words, not all of the models within the 
$\Delta \chi^2 = \bar{\chi}^2_{\mbox{\scriptsize min}}$ range around the 
best-fit model are consistent with the general result obtained by Keeton et al.\ 
(1998), and this in turn implies that the derived range of the Hubble constant 
would be narrower than obtained above if we required the model position angle to be
in agreement with the light position angle. [If a quadrupole term were used 
to account for the cluster contribution, PA$_{\mbox{\scriptsize model}}$ would 
be $\sim -40\deg$ (or, $140\deg$) implying that the dark matter be oriented 
nearly orthogonal to the light in contrast to our best-fit model and
the general result of Keeton et al.\ (1998).] 
The predicted position of the cluster's mass center off from the galaxy 
is in the northeast direction consistent with the Fischer et al.\ (1997) 
determination and the prediction of the FGSE+SIS model by Ba99. The PEM+PS model 
predicts that the cluster's center of mass is closer to the galaxy than, and the 
cluster is less massive than Fischer et al.\ (1997) found. 

Since the lowest $\bar{\chi}^2$ of the PEM+PS model is not very good, we are 
led to consider a possibility of elliptical mass distribution for the cluster.
We find that $\chi^2_{\mbox{\scriptsize min}}$ decreases slightly if an 
ellipticity is given to the cluster, but 
$\bar{\chi}^2_{\mbox{\scriptsize min}}$ worsens. 
Moreover, this has little effect on the derived value of $H_0$. 
The result of the fit shows that the cluster would be oriented
to the north if it is elliptical (i.e.\ PA$_{\mbox{\scriptsize cluster}}$
$\sim$ $0\deg$). Finally, we consider allowing the lensing galaxy's positions 
to be free parameters. In this case, the observed light distribution is 
completely ignored. In particular, it is assumed that the center of mass could
be very different from the position of the centroid of the light. We find that 
the lowest $\bar{\chi}^2$ improves to 
$\bar{\chi}^2_{\mbox{\scriptsize min}}\approx 4.7$ 
($\chi^2_{\mbox{\scriptsize min}} \approx 28.$), 
but the derived value of $H_0$ is unaffected if the lensing galaxy's positions 
are free parameters. The modeled lens positions relative to the VLBI positions
of G$'$ are $\Delta$RA = 28 mas and $\Delta$Dec = $-97$ mas.

\section{Discussion}
Recent observations of the lensing galaxy and cluster of Q0957+561 (Be97; 
Fischer et al.\ 1997) motivated modeling the galaxy by an 
elliptical mass distribution and modeling the cluster by a power-law sphere 
with an extended core. The PEM+PS model (\S3) was fitted to both the VLBI 
constraints (Ba99) and the HST constraints (Be97) of the system to derive 
a value for $H_0$. The best-fit model has
$\bar{\chi}^2_{\mbox{\scriptsize min}} \approx 6$ which is better than the
best-fits achievable by previous models. Here it should be emphasized that 
the relative improvement of the fit was achieved by using an elliptical mass 
distribution for the galaxy {\it and} using a mass sphere for the cluster, 
replacing the quadrupole term. The only other model which gives 
a fit of comparable quality is the FGSE model of Ba99, where the galaxy is 
modeled by an elliptical King profile mass distribution with an extended core 
along with a large effective black hole (BH) mass at the galaxy's center, 
while the cluster is modeled using a quadrupole 
term or a singular isothermal sphere (SIS). The FGSE model suggests that the 
Hubble constant is $h = 1.23^{+0.22}_{-0.23}$ and $1.13^{+0.32}_{-0.27}$ 
($\Delta \chi^2 = 4 \bar{\chi}^2_{\mbox{\scriptsize min}}$ CLs) using the 
quadrupole term and the SIS for the cluster, respectively. 
A few comments appear to be relevant regarding the FGSE model.
First, the model requires a very large point mass of $\sim 10^{11}$ M$_{\odot}$, 
which does not seem observationally motivated or supported.
Second, the value of $H_0$ derived from the FGSE model is inconsistent with
other recent independent determinations of $H_0$, especially the determination
by Impey et al.\ (1998) from the lens PG 1115+080, who found $h \lesssim 0.70$
regardless of the choice of lens models. On the other hand,
the PEM+PS model predicts that $h = 0.69^{+0.18}_{-0.12} (1-\kappa)$ and 
$0.74^{+0.18}_{-0.17} (1-\kappa)$ 
($\Delta \chi^2 = \bar{\chi}^2_{\mbox{\scriptsize min}}$ CLs) with 
the cluster's mass constrained and unconstrained, respectively. 
Since $\kappa > 0$, the PEM+PS model predicts a significantly lower
value for the Hubble constant than the FGSE model. Just recently,
Bernstein \& Fischer (1999) have considered breaking power-law index 
of the galaxy to model its potential by up to 3 subregions of 
independent elliptical mass distribution, 
and have included higher-order-than-quadrupole terms in the multipole 
expansion of the cluster potential. Although Bernstein \& Fischer's (1999)
models are more sophisticated than previous (published) models and
their models give $\chi^2 \sim$ N$_{\mbox{\scriptsize dof}}$, direct 
comparison with the above models (i.e.\ PEM+PS, FGSE) is not possible 
because they did not use the full set of magnification constraints 
(but used only the flux ratios) and de-emphasized the observed jet 
positions by using $\approx 5\sigma$ in the definition of their $\chi^2$,
and because of other differences in the fit procedure, e.g., their fixing the
redshifts of the blobs and knots at $z=1.41$, calculating the positional 
$\chi^2$ on the source plane by transforming the image position variances to 
it.\footnote{Using the same constraints used by Bernstein \& Fischer (1999),
but calculating the positional $\chi^2$ on the image plane (thus with no
approximations) and allowing the redshifts of the blobs and knots each to be 
free parameters, we find, for example, that a PEM+PS model with 
$\nu_1 = 1.86$ and $d_{12} = 4''$ have $\chi^2 \approx 10$ with 
N$_{\mbox{\scriptsize dof}} = 4$ which is comparable, in the fit quality, 
to the the best-fit models of Bernstein \& Fischer (1999).}
They have found $h = 1.04^{+0.31}_{-0.23} (1-\kappa)$ (95\% CL).

The mere values of $\bar{\chi}^2_{\mbox{\scriptsize min}}$ of the FGSE model 
and the PEM+PS model cannot distinguish between them. 
Even worse, neither of the models can be considered
to be acceptable if the adopted observational errors are true and assumed 
normally distributed, even though they give present ``best'' fits. In other
words, we are at present faced with two difficulties in determining $H_0$ from 
the Q0957+561 lens. One is the poor fit of the present best models to the 
present observational constraints. This might be, from a modeler's point 
of view, partly because the reported measurement errors were underestimated 
and/or possible systematic errors (see below) could not be taken into account. 
In fact, Ba99 point out that the radio jet positions and the magnification 
constraints, derived from the VLBI data using even the improved analysis 
procedure, should be treated with some caution. For example, superluminal 
motion over the period of the time delay or a milli-arcsecond scale deflection
by a globular cluster could have influenced the observed jet positions
by a scale much larger than their formal uncertainties of $\sim 0.1$ mas.
For this reason, it is useful to identify the individual contributions of
the observational constraints to the total $\chi^2$. Table 6 shows the
$\chi^2$ contributions of the observational constraints for the PEM+PS model.
Remarkably, the jet positions contribute most to the $\chi^2$, and
they have the second largest $\chi^2$/N$_{\mbox{\scriptsize constraint}}$
after the blob magnification ratio.\footnote{The large $\chi^2$ contribution 
of the blob magnification ratio might not be a surprising result since
current data on the blobs' surface brightnesses do not permit a very reliable
estimate of the centroid magnification ratio (Bernstein 1998, private
communication).}  This could be an indication that the above-mentioned 
effects are present. Nonetheless, the poor overall fit is an indication that 
more sophisticated and realistic models need to be considered in the future 
to better describe the mass distribution of the lens. As seen in \S 4, 
the introduction of an ellipticity to the cluster does not improve 
$\bar{\chi}^2_{\mbox{\scriptsize min}}$, at least with
the present observational constraints. 

Up until now in lens research, 
an elliptical mass density of a constant ellipticity [e.g.\ the functional
form of eq.\ (1)] has been used, as regarded relatively realistic, to model 
a lensing object. While the mass distribution of equation (1) is very flexible 
in that all of its parameters are unrestricted, it has several limitations 
in describing the true mass distribution of a lensing object, e.g., 
the elliptical lensing galaxy of Q0957+561. First, a truncation radius is 
not included in its functional form. For this reason its total mass diverges 
for a radial profile not steeper than the modified Hubble profile (i.e.\ for 
$\nu \leq 3$). Second, the ellipticity of equation (1) is constant over the 
entire range of $r$. The observed light distribution of the lensing galaxy in 
Q0957+561 (Be97) shows a varying ellipticity as a function of $r$ in the inner 
region of the galaxy. Even when dark mass is the dominant mass component, it 
would be surprising if the ellipticity of the mass distribution was constant 
over the entire range of $r$. Third, the position angle of equation (1) is 
constant over the entire range of $r$. We note that the observed position 
angle of the lensing galaxy in Q0957+561 (Be97) is scattered for a range of 
$r$. Finally, equation (1) neglects small scale perturbations or substructure
(Mao \& Schneider 1998) but assumes a smooth distribution of mass. 
Bernstein \& Fischer's (1999) models partly account for the second and third
points, however, the evaluation of their models depends on interpretation
of the current observational constraints. Realistic incorporation of the 
above-mentioned deviations from equation (1) may improve the fit unambiguously
in the future.

In this and previous studies (Ba99; Grogin \& Narayan 1996), attempts have 
been made to put constraints on $H_0$ using lens models despite the poor 
fits. We are faced with another difficulty in those attempts; two classes 
of models with similar quality fits can give disagreeing values of $H_0$, 
and the range of $H_0$ for each class of model is large. This ``degeneracy 
problem'' was also encountered in the recent efforts to determine $H_0$ 
using the lens PG 1115+080 by Impey et al.\ (1998). Their dark mass dominated 
model and constant mass-to-light ratio model give similar quality fits 
($\chi^2_{\mbox{\scriptsize min}} \approx 3$ to 4 with 
N$_{\mbox{\scriptsize dof}} = 1$), but imply $h = 0.44\pm 0.04$ and 
$h = 0.65\pm 0.05$ (for $\Omega_0 = 1$), respectively. 
Here we focus on the degeneracies in the models of Q0957+561.
Since the FGSE model requires a point mass of $\sim 10^{11}$ M$_{\odot}$
at the center of the galaxy, which appears to be rather unnatural 
(even if we interpret it as an effective 
mass rather than a physical BH mass), we would favor 
the PEM+PS model over the FGSE model unless the FGSE model was found to give
a much better fit than the PEM+PS model. At the same time, the large range of 
$h$ in the PEM+PS model can be significantly reduced if the cluster's mass 
distribution is more accurately determined from observations in the
future (e.g.\ using AXAF observations, or improved data on weak lensing 
effects), and/or the galaxy's mass is securely inferred from the measured
velocity dispersions of the galaxy using, e.g., realistic stellar dynamics
models for the PEM galaxy model. Observational studies of the two optical 
blobs and two knots (e.g.\ measurements of their redshifts and better 
astrometric and photometric data) will also be useful.

In conclusion, the PEM+PS model of Q0957+561 shows appreciable 
($\approx 2 - 3 \sigma$) discrepancies with some of the present observational 
constraints (in particular, the radio jet positions and the optical blob 
magnification ratio) while it is consistent with many others. 
For this reason, the determined value of $H_0$ should be taken
with some caution until the discrepancies are resolved.
Nevertheless, the modeled mass distributions of the galaxy and the 
cluster are consistent with the observations of the galaxy and the cluster 
(in fact, the PEM+PS model was motivated by recent observations 
of the galaxy and the cluster). This increases the likelihood that 
this lens can be used to determine $H_0$ accurately in the future 
with the aid of better data and more realistic lens models.
It is worth mentioning that our present determined value of $H_0$ is
consistent with the independent determination by Impey et al.\ (1998) using 
the lens PG 1115+080 and recent determinations from Type Ia supernovae 
observations, providing the cluster's convergence in the region of the galaxy 
of Q0957+561 is $\kappa \sim 0.1 - 0.2$ as estimated from recent
observations (Chartas et al.\ 1998; Fischer et al.\ 1997). For example, 
if we adopt $\kappa \approx 0.26\pm 0.08$  ($1\sigma$) derived by 
Bernstein \& Fischer (1999) from the Fischer et al.\ (1997) observation, 
we find \[ H_0 = 51^{+14}_{-10} \hspace{0.12in} \mbox{and} \hspace{0.12in}
  55^{+15}_{-14} \hspace{0.12in}  \mbox{km s$^{-1}$ Mpc$^{-1}$}, \]
with and without a constraint on the cluster's mass, respectively.
These values of $H_0$ are in agreement with the values derived from Type Ia 
Supernovae observations [Branch 1998 ($H_0 = 60\pm 10$ km s$^{-1}$ Mpc$^{-1}$);
Schaefer 1998 ($H_0 = 55\pm 8$ km s$^{-1}$ Mpc$^{-1}$)] as well as the values 
obtained by Impey et al.\ (1998). This could be an indication that both the
results from Q0957+561 and PG 1115+080 are on the right track. 
The coming years will be an exciting period of time for 
our observational and theoretical efforts to determine more accurately the 
Hubble constant directly from the time delays of gravitationally lensed 
systems.

\bigskip
I would like to thank Dr.\ Bernstein for comments on the HST blobs and 
providing the revised blob magnification ratio prior to publication and Dr.\
Barkana for sending a revised version of the article cited in this paper.
My special thanks are due to Dr.\ Turnshek for helpful discussions, careful 
reading of the manuscript and numerous comments, and his encouragement 
and support. Financial support from the Andrew Mellon Fellowship is 
gratefully acknowledged. The anonymous referee provided a very useful review 
of the paper, which was helpful in clarifying/improving the analysis and 
presentation.

\newpage

\centerline{\bf Table 1}
\centerline{\bf ``Primary'' Lensing Constraints$^1$}
\tabskip=4.0em
\halign to \hsize{#\hfil&\hfil#\hfil&\hfil#\hfil
\cr
\noalign{\vskip6pt\hrule\vskip3pt\hrule\vskip6pt}
Obs. Constraint  & value (uncertainty)  & reference(s)  \cr
\noalign{\vskip6pt\hrule\vskip6pt}
$\Delta\alpha(A_5-A_1)$ (mas) $^2$ & 16.6 (0.1) & 1 \cr
$\Delta\delta(A_5-A_1)$ (mas) $^2$ & 45.6 (0.1) & 1  \cr
$\Delta\alpha(B_5-B_1)$ (mas) $^2$ & 18.32 (0.07) & 1 \cr
$\Delta\delta(B_5-B_1)$ (mas) $^2$ & 55.8 (0.2) & 1 \cr
$M_1$ at $A_5$ $^3$ & 1.15 (0.03) & 1 \cr
$M_2$ at $A_5$ $^3$ & $-0.56$ (0.03) & 1 \cr
$\phi_1 (\deg)$ at $A_5$ $^3$ & 18.76 (0.04) & 1 \cr
$\phi_2 (\deg)$ at $A_5$ $^3$ & 107 (7) & 1 \cr
$M_1$ at $A_1$ $^3$ & 1.27 (0.03) & 1 \cr
$M_2$ at $A_1$ $^3$ & $-0.58$ (0.04) & 1 \cr
$\Delta\alpha(A_1-B_1)$ ($''$) & $-1.25254$ (0.00004) & 1, 2 \cr
$\Delta\delta(A_1-B_1)$ ($''$) & 6.04662 (0.00004) & 1, 2 \cr
Rel.\ Mag. ($B_1/A_1$) & 0.747 (0.015) & 3 \cr
$\Delta\alpha({\mbox G}'-B_1)$ $^4$ & 0.181 (0.001) & 4 \cr
$\Delta\delta({\mbox G}'-B_1)$ $^4$ & 1.029 (0.001) & 4 \cr
\noalign{\vskip3pt\hrule}
}

NOTE--(1) All of the positions are for J1950.0.
(2) See Table 2 for the correlation coefficients.
(3) See Table 3 for the correlation coefficients.
(4) Since the VLBI position of G$'$ is consistent with the HST position
by Bernstein et al.\ (1997) at a 10 mas level, we have chosen to use this
position. However, the model fit is unaffected if the HST position is used.

REFERENCES--(1) Barkana et al.\ (1998); (2) Falco, Gorenstein, \& Shapiro 1991;
(3) Conner, Leh\'{a}r, \& Burke 1992; (4) Gorenstein et al.\ 1983

\bigskip
\centerline{\bf Table 2}
\centerline{\bf Correlation Coefficients for Radio Jet Positions$^1$}
\tabskip=1.8em
\halign to \hsize{\hfil#\hfil&\hfil#\hfil&\hfil#\hfil&\hfil#\hfil&\hfil#\hfil
\cr
\noalign{\vskip6pt\hrule\vskip3pt\hrule\vskip6pt}
  & $\Delta\alpha(A_5-A_1)$ & $\Delta\delta(A_5-A_1)$ & $\Delta\alpha(B_5-B_1)$
  & $\Delta\delta(B_5-B_1)$ \cr
$\Delta\alpha(A_5-A_1)$ &  1.00  &     &     &      \cr
$\Delta\delta(A_5-A_1)$ &  0.26  & 1.00 &    &      \cr
$\Delta\alpha(B_5-B_1)$ & -0.40 & 0.27 & 1.00 &   \cr
$\Delta\delta(B_5-B_1)$ & -0.12 & 0.19 & 0.36 & 1.00 \cr
\noalign{\vskip3pt\hrule}
}

NOTE--(1) Adapted from Barkana et al.\ (1998)

\newpage

\centerline{\bf Table 3}
\centerline{\bf Correlation Coefficients for Magnification Constraints$^1$}
\tabskip=1.2em
\halign to \hsize{\hfil#\hfil&\hfil#\hfil&\hfil#\hfil&\hfil#\hfil&\hfil#
\hfil&\hfil#\hfil&\hfil#\hfil\cr
\noalign{\vskip6pt\hrule\vskip3pt\hrule\vskip6pt}
  & $M_1$ at $A_5$  & $M_2$ at $A_5$ & $\phi_1$ at $A_5$ & $\phi_2$ at $A_5$ 
 & $M_1$ at $A_1$ & $M_2$ at $A_1$ \cr
$M_1$ at $A_5$  & 1.00 &   &   &   &   &   \cr
$M_2$ at $A_5$ & 0.05 & 1.00 &   &   &   &   \cr
$\phi_1$ at $A_5$ & -0.27 & 0.25 & 1.00 &   &   &   \cr
$\phi_2$ at $A_5$  & -0.34 & -0.74 & -0.16 & 1.00 &  &   \cr
$M_1$ at $A_1$ & -0.95 &  -0.02 & 0.20 & 0.16 & 1.00 &   \cr
$M_2$ at $A_1$ & -0.06 & 0.985 & 0.31 & -0.70 & 0.09 & 1.00 \cr
\noalign{\vskip3pt\hrule}
}

NOTE--(1) Adapted from Barkana et al.\ (1998)

\bigskip

\centerline{\bf Table 4}
\centerline{\bf ``Secondary'' Lensing Constraints$^{1,2}$}
\tabskip=4.0em
\halign to \hsize{#\hfil&\hfil#\hfil&\hfil#\hfil
\cr
\noalign{\vskip6pt\hrule\vskip3pt\hrule\vskip6pt}
Obs. Constraint  & value (uncertainty)  & reference  \cr
\noalign{\vskip6pt\hrule\vskip6pt}
$\Delta\alpha$ (Blob 2) ($''$)  & 1.72 (0.05) &  1   \cr
$\Delta\delta$ (Blob 2) ($''$)  & 0.98 (0.05) &  1   \cr
$\Delta\alpha$ (Blob 3) ($''$)  & $-2.70$ (0.05) &  1   \cr
$\Delta\delta$ (Blob 3) ($''$)  & 4.48 (0.05) &  1   \cr
Rel.\ Mag.\ (Blob 2/Blob 3) & 0.41 ($^{+0.14}_{-0.11}$) & 2 \cr
$\Delta\alpha$ (Knot 1) ($''$)  & 0.13 (0.05) & 1 \cr
$\Delta\delta$ (Knot 1) ($''$)  & $-1.53$ (0.05) & 1 \cr
$\Delta\alpha$ (Knot 2) ($''$)  & $-0.29$ (0.05) & 1 \cr
$\Delta\delta$ (Knot 2) ($''$)  & $-1.41$ (0.05) & 1 \cr
Rel.\ Mag.\ (Knot 1/Knot 2) & 0.76 ($^{+0.68}_{-0.36}$) & 1 \cr
\noalign{\vskip3pt\hrule}
}

NOTE--(1) The redshifts of the objects are unknown at present.
(2) All of the positions are relative to the core of the quasar B
(e.g.\ component $B_1$) and for J1950.0.

REFERENCE--(1) Bernstein et al.\ (1997); 
(2) Bernstein (1998, private communication)

\newpage

\centerline{\bf Table 5}
\centerline{\bf Fitted Values of the Model Parameters$^1$}
\tabskip=4.0em
\halign to \hsize{#\hfil&\hfil#\hfil&\hfil#\hfil
\cr
\noalign{\vskip6pt\hrule\vskip3pt\hrule\vskip6pt}
Parameter  & Best-fit value & 
$\Delta \chi^2 = \bar{\chi}^2_{\mbox{\scriptsize min}}$ range \cr
\noalign{\vskip6pt\hrule\vskip6pt}
$\nu_1$   &  1.72  &  $1.62 \lesssim \nu_1  \lesssim 1.80$ \cr
$r_1$ ($h^{-1}$ kpc) & 10$^{-4}$ (fixed) &  --- \cr
$\Sigma_1$ ($10^{10}$ M$_{\odot}$ kpc$^{-2}$) & 723.4  & 
   $284.5 \lesssim \Sigma_1 \lesssim 1754.$  \cr
$\epsilon_1$      &  0.184  &  $0.007 \lesssim \epsilon_1 \lesssim 0.377$ \cr
$\theta_1$ ($\deg$) & 64.0 &   $5.7  \lesssim  \theta_1 \lesssim 166.0$ \cr
$\nu_2$   & 2 (fixed)  &  ---  \cr
$r_2$ ($h^{-1}$ kpc)  &  15 (fixed) & --- \cr
$\Sigma_2$ ($10^{10}$ M$_{\odot}$ kpc$^{-2}$) & 0.246 & 
              $0.164 \lesssim \Sigma_2 \lesssim 0.389$ \cr
$d_{12}$ ($''$) &  9.   & $5. \lesssim  d_{12}  \lesssim 13.$ \cr
PA$_{12}$ ($\deg$) & 51.8 & $38.0 \lesssim \mbox{PA}_{12} \lesssim 60.0$ \cr
$z_{\mbox{\scriptsize blob}}$ & 1.43 &  $1.41 \lesssim 
    z_{\mbox{\scriptsize blob}}  \lesssim 1.47$ \cr
$z_{\mbox{\scriptsize knot}}$ & 1.34 & $1.22 \lesssim 
    z_{\mbox{\scriptsize knot}}  \lesssim 1.54$ \cr
\noalign{\vskip3pt\hrule}
}

NOTE--(1) The source positions of the core, the jet, the blob, and the knot
are not included here. The total number of {\it free} parameters including the
eight source positions is 17.

\bigskip

\centerline{\bf Table 6}
\centerline{\bf Individual $\chi^2$ Contributions from
 the Observational Constraints}
\tabskip=3.5em
\halign to \hsize{#\hfil&\hfil#\hfil&\hfil#&\hfil#\hfil
\cr
\noalign{\vskip6pt\hrule\vskip3pt\hrule\vskip6pt}
Constraints & N$_{\mbox{\scriptsize constraint}}$ & $\chi^2$  & 
($\chi^2$/N$_{\mbox{\scriptsize constraint}}$) \cr
\noalign{\vskip6pt\hrule\vskip6pt}
Jet positions & 4  & 19.00 & (4.75) \cr
Core positions & 2 &  0.00 & (0.00) \cr
G1 positions  & 2  &  0.05 & (0.03) \cr
Magnification constraints & 6 & 18.65 & (3.11) \cr
Core magnification ratio & 1 & 0.08 & (0.08) \cr
``Blob 2'' positions & 2  & 3.24 & (1.62) \cr
``Blob 3'' positions & 2  & 0.18 & (0.09) \cr
Blob magnification ratio & 1 & 8.26 & (8.26) \cr
``Knot 1'' positions & 2 & 0.00 & (0.00) \cr
``Knot 2'' positions & 2 & 0.00 & (0.00) \cr
Knot magnification ratio & 1 & 0.19 & (0.19) \cr
\noalign{\vskip6pt\hrule\vskip6pt}
Total & 25 & 49.6  \cr
\noalign{\vskip3pt\hrule}
}

\newpage
\begin{figure}
\centerline{\psfig{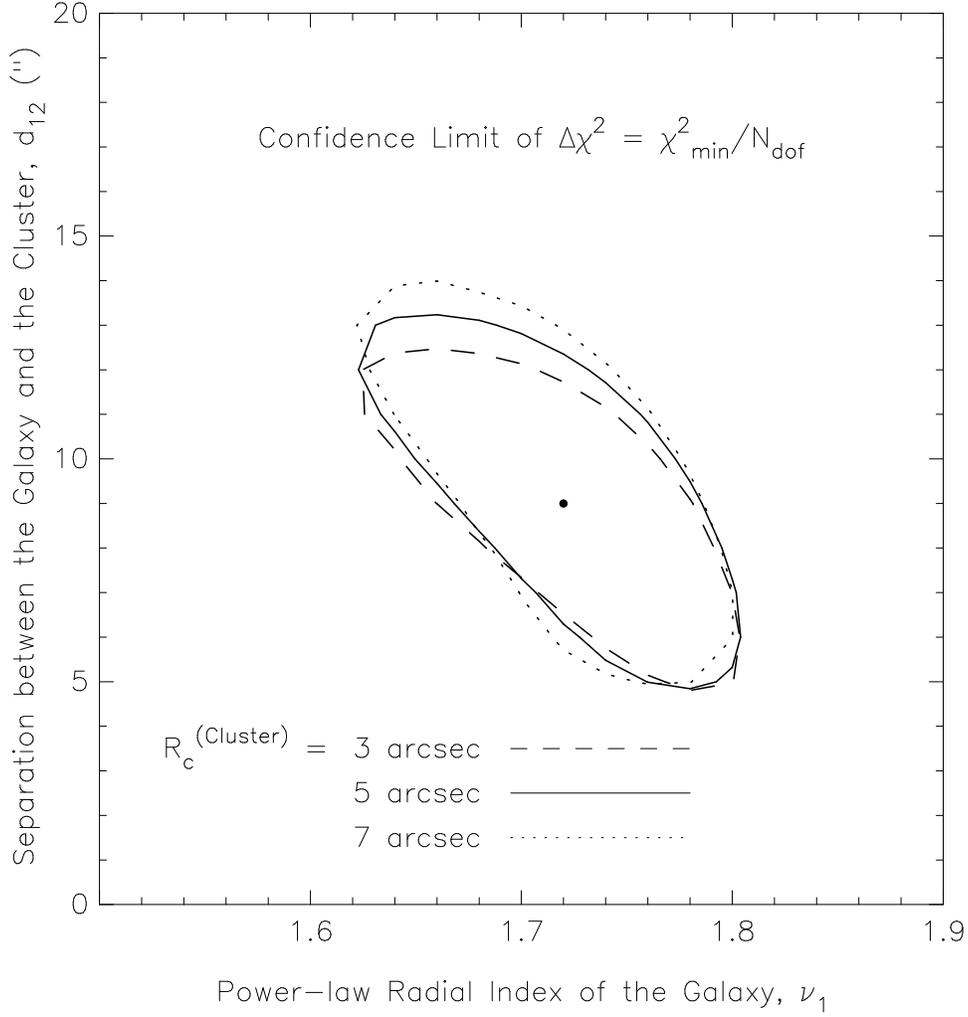}}
\caption[]{A confidence region of parameter space defined by 
$\Delta \chi^2 = \bar{\chi}^2_{\mbox{\scriptsize min}}$ (see \S4) around 
the best-fit model ($\nu_1 = 1.72$ and $d_{12} = 9''$). 
The two-dimensional parameter space is spanned by the radial 
index of the lensing galaxy ($\nu_1$) and the (projected) separation between 
the galaxy and the cluster ($d_{12}$). The cluster is represented by a
power-law mass sphere with a non-singular core in the basic model (eq.\ [2]), 
and here a fixed value of $\nu_2 = 2$ (``isothermal'' distribution) is used.
To illustrate how the confidence region is affected when the value of the 
cluster's core radius is varied from the best-estimated value of $5''$ 
(Fischer et al.\ 1997), core radii of $3''$ and $7''$ are considered as well
as the best-estimated value. We find that the confidence region is altered 
only slightly as the core radius is varied.}
\end{figure}
 
\newpage
\begin{figure}
\centerline{\psfig{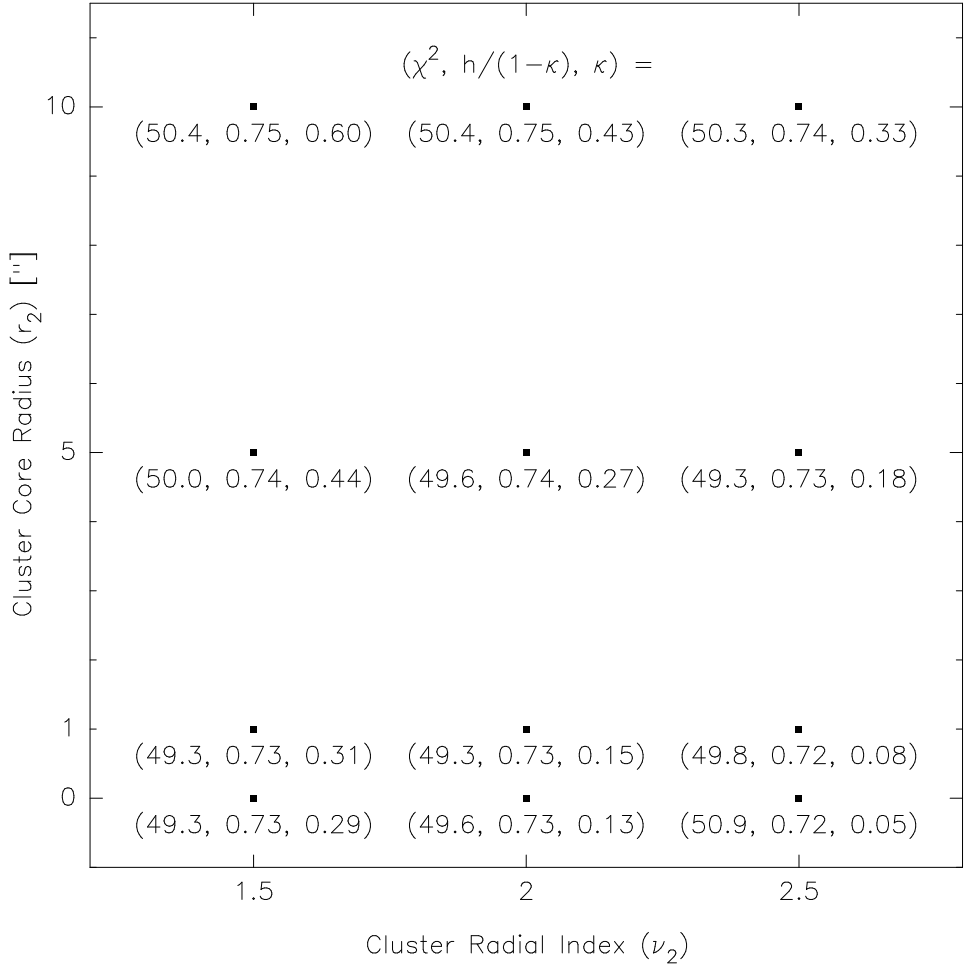}}
\caption[]{Degeneracy of cluster parameters and effect on the derived 
value of $h$. We consider a range of cluster parameters, 
$1.5 \leq \nu_2 \leq 2.5$ and $0 \lesssim r_2 \lesssim 10''$, 
for the fixed galaxy radial index $\nu_1 = 1.72$ and galaxy-cluster 
separation $d_{12} = 9''$. We see that $\Delta\chi^2 \lesssim 1.6$ and 
$\Delta [h/(1-\kappa)] \lesssim 4\%$ as $\kappa$ ranges from $0.05 - 0.60$ 
within the parameter space. This means that as we change either the cluster's 
radial index or core radius from the observationally favored values of 
$\nu_2 = 2$ and $r_2 \approx 5''$ (Fischer et al.\ 1997), $\Delta\chi^2$ is 
insignificant and the scaling $h \propto (1-\kappa)$ holds to an error of 
4\% (see \S 4).}
\end{figure}
 
\newpage
\begin{figure}
\centerline{\psfig{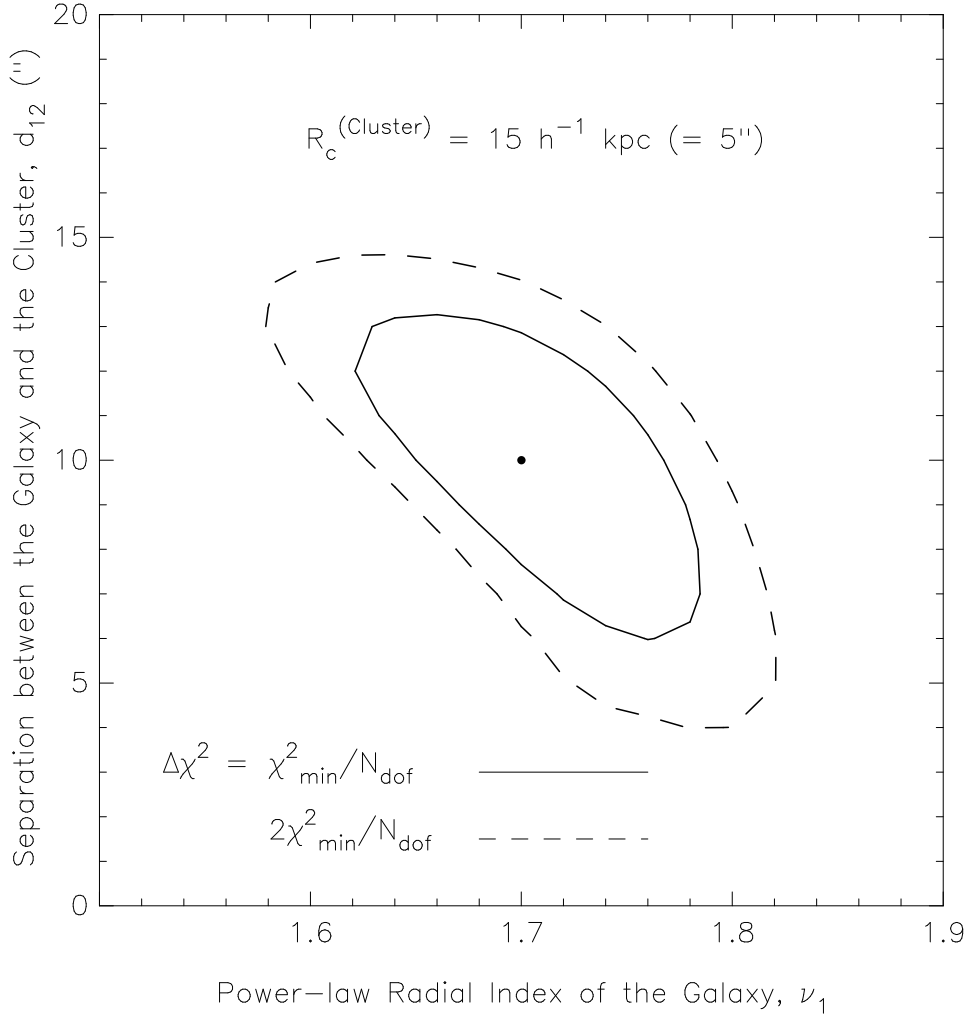}}
\caption[]{Confidence regions, in the same parameter space as in Figure 1,
of the model with the cluster core radius of $5''$. Here the cluster's mass 
is constrained using the observationally derived value by Fischer et al.\ 1997.
The parameter space is slightly better constrained compared with the case
(Fig.\ 1) when the cluster's mass is not used as an observational constraint.
In fact, if the cluster's mass were determined more accurately by 
observations, the confidence region of parameter space could be reduced 
significantly, thereby reducing the allowed range of $H_0$.}
\end{figure}
 
\newpage
\begin{figure}
\centerline{\psfig{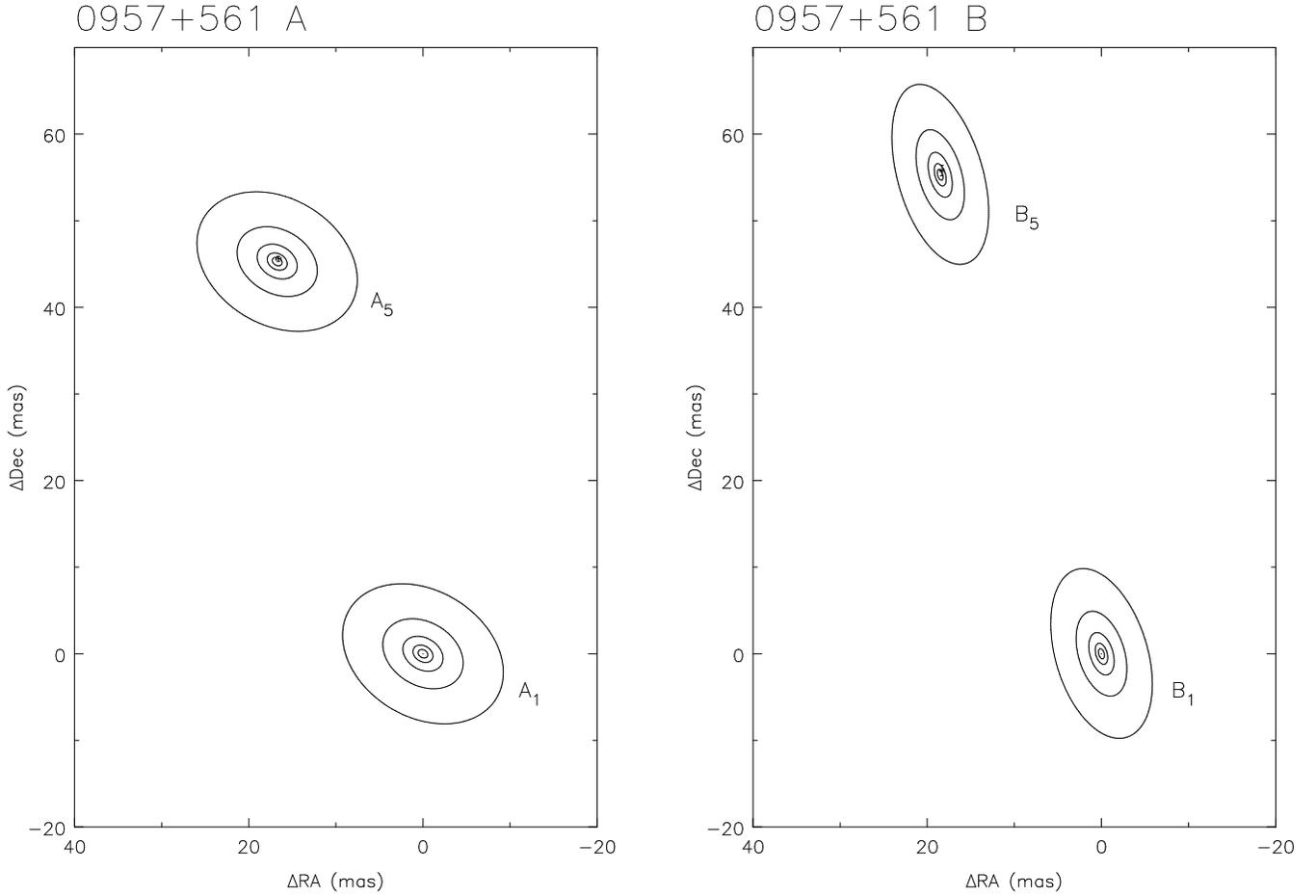}}
\caption[]{Theoretical images of the radio core and the brightest radio 
jet of Q0957+561 for the best-fit model (see Table 5). The five contour levels
for the core and the jet correspond to 1, 2, 4, 8, and 16 pc from their 
respective centers on the source plane. The observed jet positions relative to
their respective cores are also shown. The error bars represent $3\sigma$ 
uncertainties. North is on the top and east is to the left.}
\end{figure}
 
\newpage
\begin{figure}
\centerline{\psfig{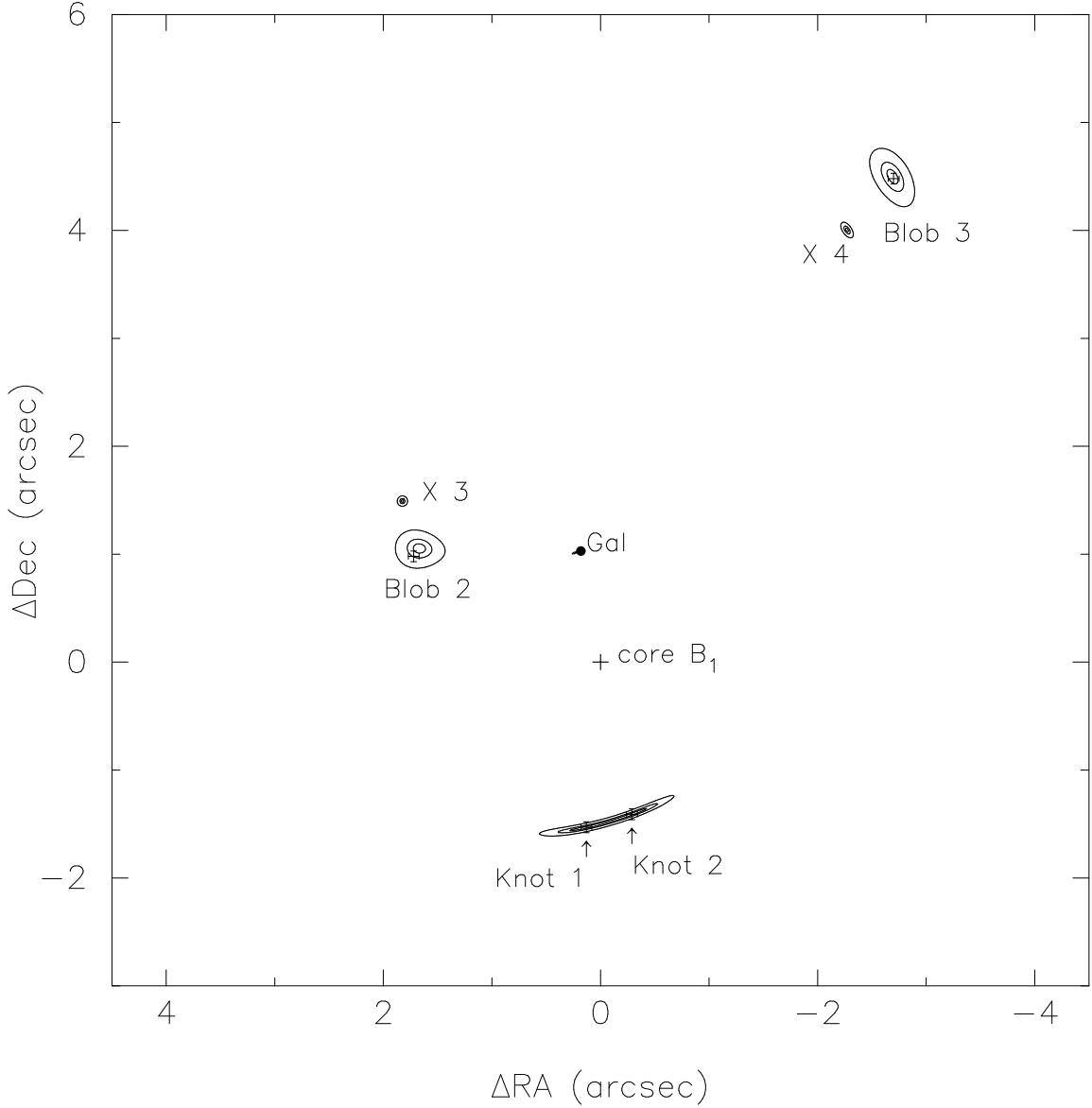}}
\caption[]{Theoretical images of the optical blob and knot of Q0957+561
for the best-fit model (see Table 5). 
The three contour levels for the blob correspond to 100, 200, and 400 pc at
its estimated redshift of $z = 1.43$. The three contour levels for the knot
correspond to 25, 50, and 100 pc at its estimated redshift of $z = 1.34$.
The observed positions of the blobs and knots (Table 4) are also shown with 
$1\sigma$ error bars. The objects marked ``X 3'' and ``X 4'' are two weak
additional images of the knot predicted by the lens model. North is on the 
top and east is to the left.}
\end{figure}
 
\end{document}